\documentclass[]{emulateapj}
\begin{document}
\title{The Radio - 2 mm Spectral Index of the Crab Nebula Measured with 
GISMO\footnotemark[*]}
\footnotetext[*]{Based on observations carried out with the IRAM 30m Telescope. 
IRAM is supported by INSU/CNRS (France), MPG (Germany) and IGN (Spain).}

\author{R. G. Arendt\altaffilmark{1,2}, 
J. V. George\altaffilmark{3}, 
J. G. Staguhn\altaffilmark{2,3,4}, 
D. J. Benford\altaffilmark{2},
M. J. Devlin\altaffilmark{5},\\
S. R. Dicker\altaffilmark{5},
D. J. Fixsen\altaffilmark{2,3}, 
K. D. Irwin\altaffilmark{6},
C. A. Jhabvala\altaffilmark{7},
P. M. Korngut\altaffilmark{5},
A. Kov\'acs\altaffilmark{8},\\
S. F. Maher\altaffilmark{2,9},
B. S. Mason\altaffilmark{10},
T. M. Miller\altaffilmark{7},
S. H. Moseley\altaffilmark{2}, 
S. Navarro\altaffilmark{11},
A. Sievers\altaffilmark{11},\\
J. L. Sievers\altaffilmark{12},
E. Sharp\altaffilmark{2,13},
E. J. Wollack\altaffilmark{2}
}

\altaffiltext{1}{CRESST, University of Maryland -- Baltimore County,
    Baltimore, MD 21250, USA; \email{Richard.G.Arendt@nasa.gov}}
\altaffiltext{2}{NASA Goddard Space Flight Center, Code 665, Greenbelt, MD 20771, USA}
\altaffiltext{3}{University of Maryland -- College Park, College Park, MD 20742, USA}
\altaffiltext{4}{The Henry A. Rowland Department of Physics and Astronomy, Johns Hopkins University, 3400 N. Charles Street, Baltimore, MD 21218, USA}
\altaffiltext{5}{University of Pennsylvania, Physics and Astronomy, 209 South 33rd Street, Philadelphia, PA 19104, USA}
\altaffiltext{6}{NIST Quantum Devices Group, 325 Broadway Mailcode 817.03, Boulder, CO 80305, USA}
\altaffiltext{7}{NASA Goddard Space Flight Center, Code 553, Greenbelt, MD 20771, USA}
\altaffiltext{8}{University of Minnesota, 116 Church St SE, Minneapolis, MN 55414, USA}
\altaffiltext{9}{Science Systems and Applications, Inc.}
\altaffiltext{10}{National Radio Astronomy Observatory, 520 Edgemont Road, Charlottesville, VA 22903, USA}
\altaffiltext{11}{Instituto Radioastronomia Milimetrica (IRAM), Av. Divina Pastora 7, Nucleo Central, 18012 Granada, Spain}
\altaffiltext{12}{Canadian Institute for Theoretical Astrophysics, University of Toronto, ON M5S 3H8, Canada}
\altaffiltext{13}{Global Science \& Technology, Inc.}

\shorttitle{Radio - 2 mm Spectral Index of the Crab Nebula}
\shortauthors{Arendt et al.}

\begin{abstract}
We present results of 2~mm observations of the Crab Nebula, obtained using the 
Goddard-IRAM Superconducting 2~Millimeter Observer (GISMO)
bolometer camera on the IRAM 30~m telescope. Additional 3.3 mm observations with the
MUSTANG bolometer array on the Green Bank Telescope are also presented.
The integrated 2~mm flux density of the Crab Nebula provides no evidence for the 
emergence of a second synchrotron component that has been proposed. 
It is consistent with the radio power law spectrum, extrapolated
up to a break frequency of $\log(\nu_{b} [\rm{GHz}]) = 2.84 \pm 0.29$ or $\nu_{b} = 695^{+651}_{-336}$ GHz. 
The Crab Nebula is well-resolved by the $\sim16\farcs7$ beam (FWHM) of GISMO. Comparison to radio data at 
comparable spatial resolution enables us to confirm significant spatial variation of the spectral index 
between 21~cm and 2~mm. The main effect is a spectral flattening in the inner region of the Crab Nebula, 
correlated with the toroidal structure at the center of the nebula that is prominent in the near--IR through X-ray regime. 
\end{abstract}

\section{Introduction}
\label{sec:intro}
The Crab Nebula is the prototypical example of a pulsar wind 
nebula (PWN), plerion, or filled--center supernova remnant 
\citep{davidson,gaensler,hester2008}. As one of the brighter 
objects in the sky across the entire electromagnetic spectrum from
the radio to gamma rays, the Crab Nebula is often one of the first 
targets observed by new instruments, whether they provide new spectral windows, or offer
 improvements in sensitivity, or spatial or spectral resolution
\citep[e.g.][]{bolton,marsden,temim,waller,bowyer,fazio72}.

At radio wavelengths the Crab Nebula's integrated emission is 
well known to exhibit a power law spectrum $S_{\nu} = S_0\, \nu^{\alpha}$, with a 
spectral index $\alpha = -0.299$ \citep{baars}. 
This spectral index is fairly typical of the synchrotron emission
exhibited by PWNe, and is distinctly flatter than the spectral indices of more common
shell-like SNRs where the relativistic electrons are accelerated in the shocks of
the expanding blast wave rather than by a pulsar wind. In the optical regime, the 
synchrotron emission exhibits a steeper spectral index \citep[e.g.][]{veron}. 
Thus, a spectral break has been inferred to lie at wavelengths between 10 and 
1000 $\micron$ \citep{marsden, woltjer}. However, 
the sharpness of the break and the exact location are somewhat obscured by 
the difficulty in obtaining high precision measurements at mm wavelengths, and
the presence of stronger thermal emission from dust at mid IR wavelengths 
\citep{marsden, mezger, strom}. The dust is associated with the knots and filaments
of ejecta in the Crab Nebula, although spectral observations have not revealed
characteristic features such as silicate or PAH bands in the spectrum 
\citep{douvion, temim}.

Resolved imaging of the Crab Nebula indicates that there is very little variation 
in the radio spectral index across the Crab Nebula \citep{bietenholz1997}. 
There has been some indication of a spectral index variation between 
the radio and mm regimes, suggesting the possibility of a physically distinct, second
synchrotron component \citep{bandi}, although \cite{green} argue that the variations
may in fact be due to known temporal variability of the SNR.

In this paper, we further probe the synchrotron component using a map we obtained with the GISMO 2~mm camera on the IRAM 30 m telescope. The paper is organized as follows: in section \ref{sec:obs} we introduce the various maps that were used. In the process we describe the GISMO instrument, the data reduction package (CRUSH) and the data analysis. The 21~cm and 6~cm radio maps used for comparison are introduced. In section \ref{sec:iflux}, we study the integrated radio/sub-millimeter spectrum of the SNR. We derive the location of the break in the spectrum. In section \ref{sec:specmaps}, we present and discuss spectral index maps created by comparison of the GISMO data with the radio maps. We also discuss the correlation of features in the spectral index maps with features at other wavelengths (section \ref{sec:discussion}). Finally our conclusions are stated in section \ref{sec:concl}.

\section{Observations}
\label{sec:obs}
\subsection{GISMO Data}
\label{subsec:gismod}
The Crab Nebula was observed at the 30~m telescope located on Pico Veleta (near Granada, Spain) and operated by the Institute de Radioastronomie Millim\'etrique (IRAM) \citep{iram}. The observations were centered at 
$(\alpha_{J2000},\delta_{J2000}) = (5^{h}37^{m}45^{s}, 22\arcdeg 1' 0'')$. The data were obtained during the first run of the Goddard-IRAM Superconducting 2~Millimeter Observer (GISMO) in November of 2007.

The GISMO instrument is a bolometric camera developed for operation in the 2~mm atmospheric window. The central frequency of the GISMO band is 150 GHz, with 
$\delta \nu/\nu = 0.15$ (or a bandwidth of 22 GHz @ 150 GHz). The instrument uses an 8$\times$16 planar array of multiplexed superconducting transition edge sensor (TES) bolometers which incorporates the Backshort Under Grid (BUG) architecture \citep{allen2006}. The size and sensitivity of the detector array \citep{ref_stag} give this instrument significantly greater mapping speed at higher angular resolution in this wavelength than has previously been possible. GISMO has a pixel size of 2~mm which provides for an angular separation of $14''$ on the sky corresponding to a sampling of $0.9 \lambda/D$ at 2~mm wavelength. With an array of $8 \times 16$ pixels, the GISMO instrument has a gapless field of view of $2'\times4'$ which allows it to take advantage of a significant fraction of the telescope's optical capabilities. The FWHM of the beam at 2~mm is $\sim16\farcs7$, which is $\sim15\%$ 
wider than the ideal diffraction limited beam.
 \cite{stag} gives a more detailed description of the instrument. On this first run of the instrument, there was a mechanical problem that caused one quadrant of the pixels to malfunction. The data reduction also identified pixels that exhibited increased noise 
and these too were excised, leaving 62 detector pixels used for this analysis.

\begin{figure}[t]
 \begin{center}
  \includegraphics[]{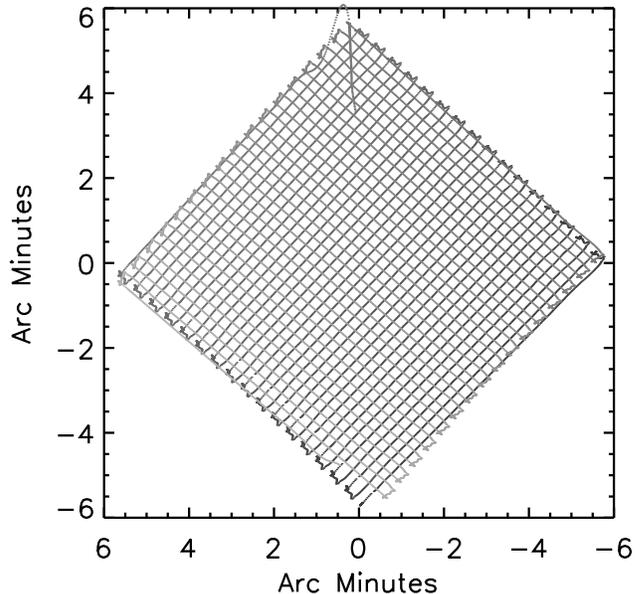}
  \caption{The scan pattern for one of the five pairs of scans used on the Crab Nebula. 
  An on-the-fly (OTF)
  scan map is executed in azimuth, immediately followed by another OTF map in elevation.
  The pattern of these scans, when projected into equatorial coordinates 
  as shown here, rotates over time as the
  parallactic angle of the target changes. The parallactic angle for this pair of scans 
  is $\sim45\arcdeg$.}
  \label{fig:scan}
 \end{center}
\end{figure}

The maps were obtained on November 12, 2007 using five pairs of on-the-fly scans. The first scan in each pair used 33 scan lines in the azimuth, $630''$ long, offset by $18''$. The second scan of each pair transposes the same pattern with 33 scan lines in elevation (Fig. \ref{fig:scan}). Each scan lasts 512 seconds, and a pair of scans requires 1035 seconds. The integration time per $3''$ pixel in the final map is $\sim$7.4 sec for each pair of scans. During a complete raster scan each element of the bolometer array then samples the whole of the Crab which has an apparent size of $420'' \times 290''$ \citep{trimble}.  

\begin{figure}[t]
\begin{center}
  \includegraphics[]{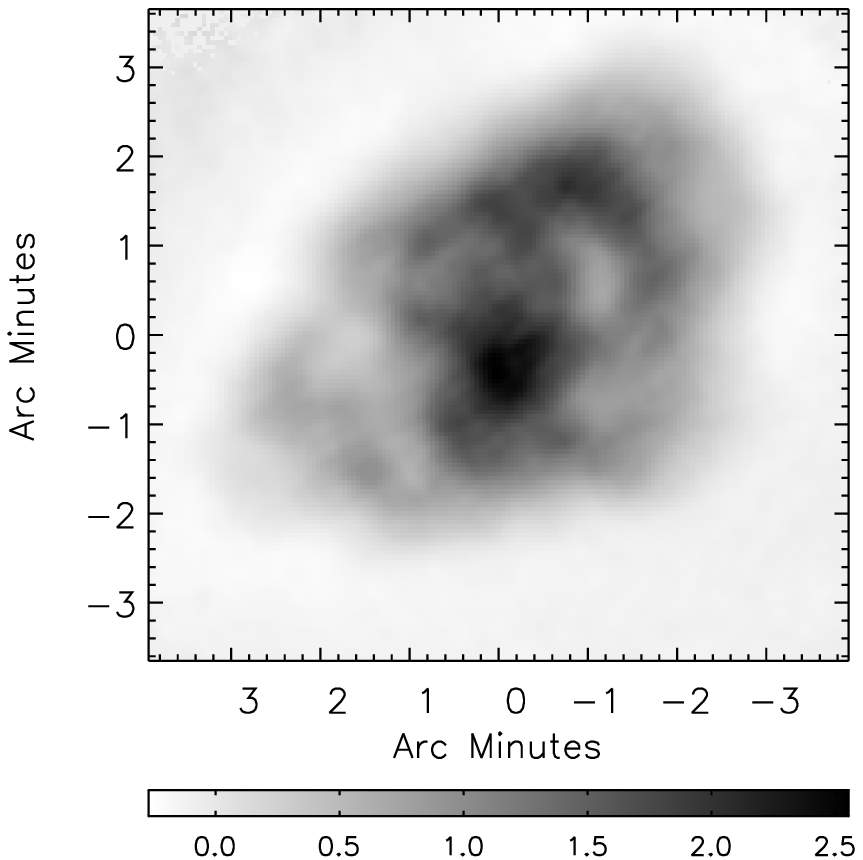}
  \includegraphics[]{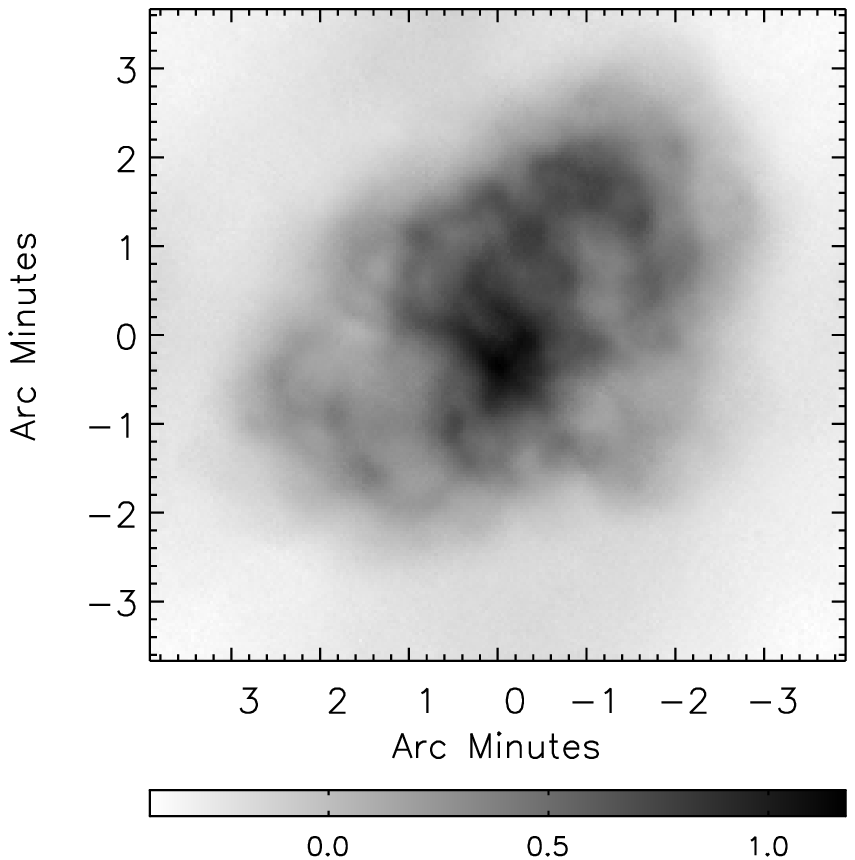}
  \caption{(top) GISMO map of the Crab Nebula at 2.0~mm. 
  The image is slightly smoothed from the intrinsic $16\farcs7$ (FWHM) resolution.
  Across the region of the Crab Nebula, the uncertainty in the intensity is 
  $0.008 < \sigma < 0.016$ Jy beam$^{-1}$.
  (bottom) MUSTANG map of the Crab Nebula at 3.3~mm. 
  The beam is $9''$ FWHM. The gray scale bars are labeled in Jy beam$^{-1}$.
  The axes indicate right ascension and declination offsets from the pulsar location: 
  $(\alpha_{\rm J2000},\delta_{\rm J2000}) = (05^h\ 34^m\ 32^s,+22\arcdeg\ 00'\ 52'')$. 
  \label{fig:images}}
\end{center}
\end{figure}

The data were reduced using the Comprehensive Reduction Utility for SHARC-2 \citep[CRUSH;][]{attila}. CRUSH was initially developed for data reduction of SHARC-2 observations, but it has been enhanced to reduce data from other instruments, including GISMO. CRUSH uses an iterated sequence of statistical estimators to separate source, atmosphere and instrument signals. As part of this reduction, the final map is smoothed
by a Gaussian function with 2/3 beam width (the CRUSH default for faint sources) to have an effective FWHM of
$\sim20''$. This smoothing is done to minimize spurious high spatial frequency features that are related to the relatively coarse sampling of the beam with the the GISMO detector pixels \citep{attila}. The final map produced by CRUSH has a pixel scale of $3''$ (Figure \ref{fig:images}). The noise level of the map varies as a function of position due to 
the coverage of the scans, but over the region where the Crab Nebula is brighter 
than 0.5 Jy beam$^{-1}$ the noise level is $\sigma < 0.016$ Jy beam$^{-1}$.

The total integrated flux density of the Crab Nebula is determined to be
$244\pm24$~Jy at 2~mm, after adjustment 
for the slightly negative local background level in the image produced by CRUSH.
This value is based on use of Saturn as a calibrator,
which was observed after the Crab Nebula during a period of poorer weather 
($\tau_{150GHz} \sim 0.3$ vs. $\tau_{150GHz} \sim 0.12$).

\subsection{MUSTANG Data}
  Additional observations of the Crab Nebula were carried out with Multiplexed Squid 
TES Array at Ninety GHz \citep[MUSTANG;][]{mustang} instrument on the Green Bank
Telescope (GBT). The data were acquired during Feb 2008, under 
project code AGBT08A$\_$056. 
The data were calibrated relative to Mars using \cite{weiland} and \cite{wright2007}.
A map was made with a maximum likelihood imaging pipeline developed for the
Atacama Cosmology Telescope 
\citep[ACT;][]{fowler} that was adapted for use with MUSTANG (Figure \ref{fig:images}).
The MUSTANG camera has $9''$ (FWHM) spatial resolution, compared to $16\farcs7$ 
for the GISMO instrument. 
Due to the comparatively small instantaneous FOV of MUSTANG ($40''\times40''$), 
the total integrated flux density for the Crab Nebula is not well constrained. 
Structures on angular scales from $4''$ to $1\farcm5$ are recovered with good fidelity. 

\subsection{Radio Data}
\label{subsec:radiod}
To understand the synchrotron emission from the Crab Nebula, 
we compared the GISMO 2~mm map to radio maps at lower frequencies.
Radio maps at two different frequencies and two different epochs were used for 
comparison. The first map at 1.41 GHz (21 cm) was obtained using all 4 configurations 
of the VLA over the period from 1987--1988 \citep{bietenholz1990}. This 
image has a cleaned beam size of $1\farcs8 \times 2\farcs0$. A second VLA map at 
5 GHz (6 cm) was obtained during 2001, and has a cleaned beam size of $1\farcs4$ 
\citep{bietenholz2001}. The 5 GHz image (only) has been corrected for the 
VLA's primary beam response ($\sim9'$ FWHM at 5 GHz).

For comparison with the GISMO image, the radio images were rescaled in both size
and brightness. We have used 1987.49 as the epoch of the 1.4~GHz image, 
and 2007.87 is the epoch for the GISMO image. This 20.38~yr time difference results
in a 3.1\% increase in the size the nebula according the the expansion measured by
\cite{bietenholz1991exp}. This expansion corresponds to $\sim0.5$ of the GISMO
beam width at the outermost edges of the SNR. 
There is a 3.5\% decrease in the radio brightness between the dates of the 1.41~GHz VLA 
and 2 mm GISMO observations, assuming a fading rate of -0.167\% yr$^{-1}$ \citep{aller}.
More recent reports \citep{vinyaikin, weiland} confirm that the fading rate is not
a strong function of frequency or time.
After applying these corrections to the radio image,
the map was convolved with a Gaussian function to match the $20\farcs$07 resolution 
of the GISMO data. Equivalent adjustments were made to the 5~GHz VLA image, resulting 
in a size increase of 1.05\% and a brightness decrease of 1.169\%. 
The similarity of results derived from both the 1.41 and 5 GHz images indicates that 
the corrections for expansion are sufficiently accurate at GISMO resolution.

\section{Integrated Flux Density and Spectral Index}
\label{sec:iflux}

\begin{figure*}[t]
\begin{center}
  \includegraphics[width=6.5in]{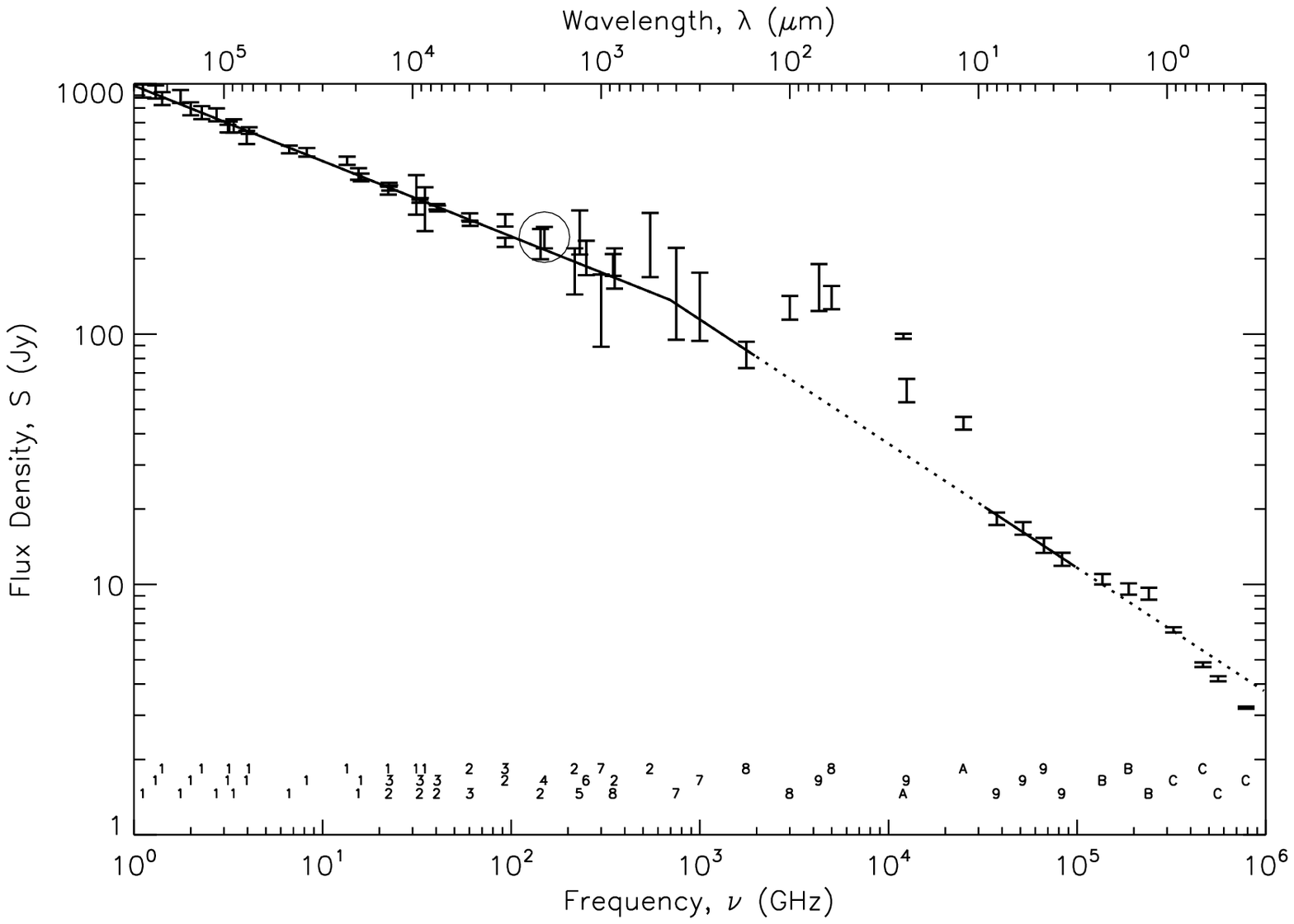}
  \caption{The integrated radio -- optical spectrum of the Crab Nebula. The 2~mm GISMO result (circled) 
  helps confirm that the break in the power law spectrum at radio wavelengths must lie a wavelengths shorter 
  than 2~mm. The line shows a broken power law fit to the data, with dotted regions indicating ranges where 
  data were ignored because of thermal emission from dust (mid--IR), or the necessity for larger extinction corrections
  (optical near--IR). Numbers along the bottom of the plot indicate the sources of the plotted data:
  (1) \cite{baars} and references therein, (2) \cite{macias-perez}, (3) \cite{weiland},
  (4) this work, (5) \cite{bandi}, (6) \cite{mezger}, (7) \cite{wright}, 
  (8) \cite{green}, (9) \cite{temim}, (A) \cite{marsden}, (B) \cite{grasd}, 
  and (C) \cite{veron}. \label{fig:spectrum}}
\end{center}
\end{figure*}

Figure \ref{fig:spectrum} shows the integrated spectrum of the Crab Nebula including the 2 mm 
GISMO flux density, and other measurements from the radio to optical wavelengths 
(30~cm -- 3000~\AA\ or 1 -- $10^6$~GHz).
All flux densities have been corrected to the GISMO epoch of 2007.87. 
The 2~mm GISMO flux density is consistent with an extrapolation 
of the 1 -- 35~GHz radio spectrum with a power law index $\alpha = -0.299\pm0.009$
as determined by \cite{baars}.

At 10~$\mu$m to 200~$\mu$m, the spectrum exhibits a bump due to thermal emission from dust.
Shorter wavelengths appear to resume a power law spectrum, but with a steeper slope than at 
the radio - mm wavelengths. We characterize the spectrum by fitting a broken power law of:
\begin{equation}
S(\nu) = \Big\{
   \begin{array}{lll}S(\nu_{b})(\nu/\nu_{b})^{\alpha_1} & & \nu < \nu_{b} \\
                     S(\nu_{b})(\nu/\nu_{b})^{\alpha_2} & & \nu > \nu_{b} \end{array}
\end{equation}
characterized by the four parameters $S(\nu_{b})$, $\nu_{b}$, $\alpha_1$, and $\alpha_2$.
The 10 - 100 $\micron$ measurements are not used to constrain the fit because of the 
possible influence of the dust emission at these wavelengths.
At the shorter wavelengths we choose to constrain the fit only with the 3.6 -- 8 $\micron$ 
{\it Spitzer} IRAC measurements of \cite{temim}, because the near-IR and optical 
measurements are much more dependent on the application of extinction corrections, and may 
also be affected by additional intrinsic steepening of the spectrum. We find
$\log[S(\nu_{b})] = 2.14\pm0.09$ (or $S(\nu_{b}) = 137^{+31}_{-25}$ Jy), 
$\log(\nu_{b}) = 2.84\pm0.29$ (or $\nu_{b} = 695^{+651}_{-336}$ GHz), 
$\alpha_1 = -0.30\pm0.01$, and 
$\alpha_2 = -0.50\pm0.03$.
The low frequency spectral index, $\alpha_1$, is consistent with \cite{baars}, while the 
high frequency spectral index, $\alpha_2$, is consistent with result 
($\alpha = -0.50\pm0.10$) obtained by \cite{douvion}
using {\it ISO} ISOCAM data covering the central $3'\times3'$ of the SNR 
at 4.5 and 11.4 $\micron$ (LW1 and LW8 filters). If the near-IR data of Graselden
were included in the fit, then $\log[S(\nu_{b})]$ would decrease by $\sim1 \sigma$, 
$\alpha_2$ would increase (flatten) by $\sim1 \sigma$, and $\alpha_1$ would remain unchanged. 
Additional inclusion of the optical measurements would result in very significant changes in 
the derived break frequency ($\log[S(\nu_{b})] ~\sim 3.85$) and a steeper high frequency 
spectral index ($\alpha_2 \sim -0.64$). However this fit systematically overshoots the IRAC 
measurements, especially at 8 $\mu$m where dust and line emission should result in an
enhancement over the synchrotron emission. Therefore we conclude that there is either a 
problem with the extinction correction or relative calibration of the optical observations,
or the intrinsic synchrotron spectrum has curvature or additional breaks such that 
it cannot be properly characterized by a single power law at mid-IR to optical wavelengths.

A detailed account of the synchrotron spectrum of the Crab Nebula 
(and other pulsar wind nebulae) is provided by the diffusive synchrotron 
radiation (DSR) model of \cite{fleishman}. In the DSR model, the magnetic field
is tangled on a range of spatial scales characterized by a power law index ($n$) and a
maximum scale length ($L$). The predicted emission spectrum will contain a break at 
a frequency that is a function of $n$ and $L$, the magnetic field strength ($B$), and 
the minimum energy ($\gamma = E/mc^2$) of the power law
distribution of relativistic particles, specifically $\nu_b \propto L^{(1-n)/(1+n)} 
B^{2/(1+n)} \gamma^2$. Adopting nominal values of these parameters
\cite{fleishman} find the break frequency to be at $\nu_b \approx 3200$ GHz.
A shift in the break frequency may be accommodated by adjusting any or all of the 
parameters. With $n$ being fairly strictly constrained to values of 
$1.54 \lesssim n \lesssim 1.6$ by the spectral index of the 
low frequency (radio) emission ($n = 1-2\alpha_1$), the value of $\nu_b$ will be most sensitive to changes
in $\gamma$ and least sensitive to changes in $L$.

The GISMO measurement and our analysis support the conclusion of \cite{green} that the 
integrated spectrum of the Crab Nebula continues as a power law from the radio regime 
down to wavelengths $<850$ $\micron$. While the \cite{bandi} 1.3 $\micron$ measurement 
lies slightly above the power law, the integrated spectrum including newer data does not exhibit a
significant excess component at millimeter wavelengths. The 170 $\micron$ ISO measurement 
\citep{green} seems to confirm the location of the break. Future observations in 
the 200 -- 800 $\micron$ range should help to better refine the location and 
sharpness of the break.

\section{Spectral Index Maps}
\label{sec:specmaps}

\begin{figure}[t]
\begin{center}
  \includegraphics[]{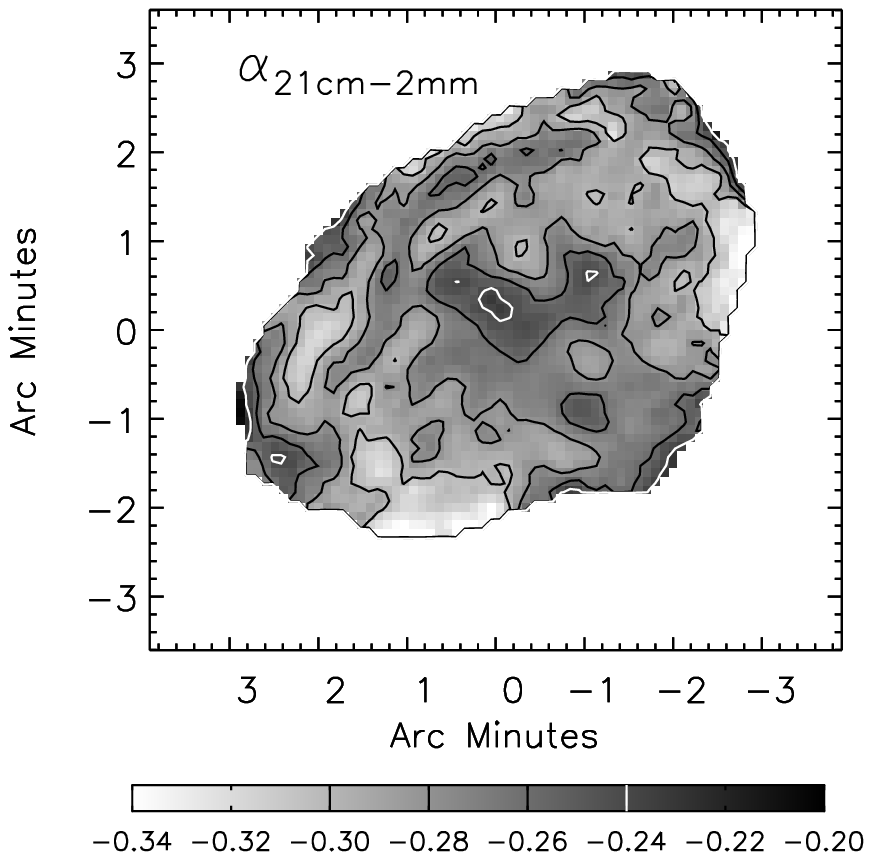}\\
  \includegraphics[]{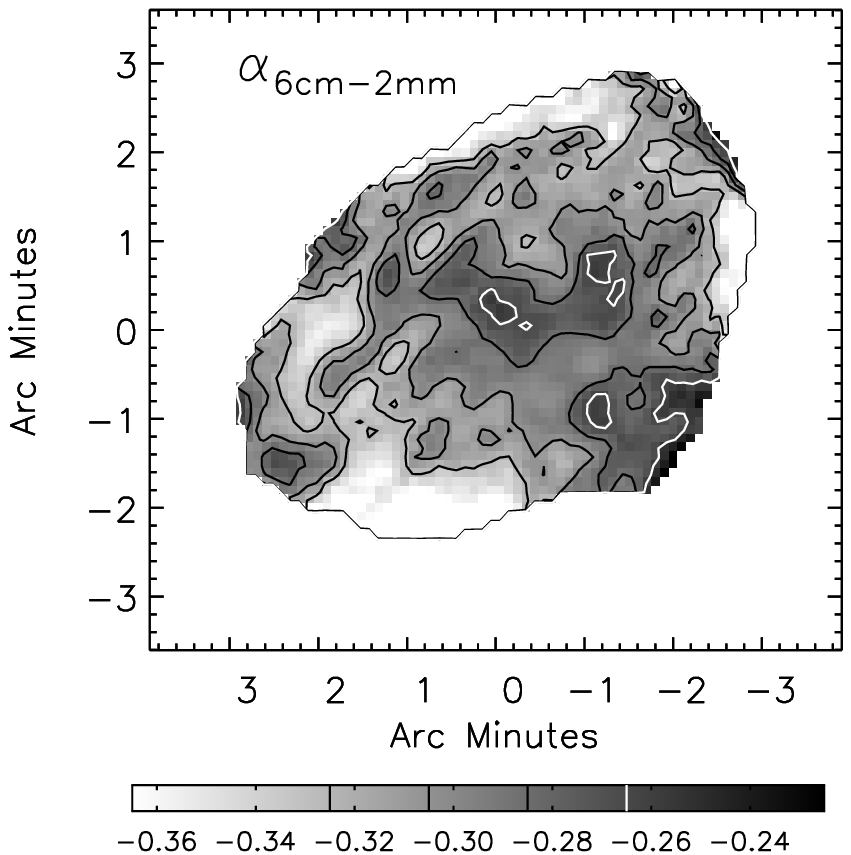}
  \caption{Radio - 2 mm spectral index maps. (top) Map of the spectral index $\alpha$, 
  for the 21~cm to 2~mm range. The contour levels are drawn at 
 $\alpha = -0.30$, -0.28, -0.26, and -0.24 (white contour).
 (bottom) Map of the spectral index $\alpha$, for the 6~cm 
  to 2~mm range. The contour levels are drawn at 
 $\alpha = -0.325$, -0.305, -0.285, and -0.265 (white contour).
 The small offset in contour levels is done to highlight the similarity in 
 structure of these two images, rather than the small shift in the mean spectral 
 index found in these comparisons to two independent radio images.
  The axes indicate right ascension and declination offsets from the pulsar location: 
  $(\alpha_{\rm J2000},\delta_{\rm J2000}) = (05^h\ 34^m\ 32^s,+22\arcdeg\ 00'\ 52'')$. 
  \label{fig:index}}
\end{center}
\end{figure}

Maps at $20''$ resolution of the Crab Nebula's spectral index between 
2~mm and radio wavelengths are shown in Figure \ref{fig:index}. These maps show significant structure
in the spectral index, and strongly resemble those of \cite{bandi} (1.3 mm - 21 cm) 
and \cite{green} (850 $\micron$ - 21 cm). There is a slight offset between the mean spectral 
index of the two maps because the radio data themselves are not an exact match 
to the $\alpha = -0.299$ spectral index. A multiplicative scale error in the absolute 
calibration of the radio (or GISMO) data will affect the mean spectral index, but 
not any relative variations in spectral index. Therefore, the contours chosen for 
Fig. \ref{fig:index} are slightly offset to emphasize the strong similarity in 
structure between the two spectral index maps.

The random uncertainties (noise) in the intensity maps 
propagate into uncertainties in the derived
spectral index. For $\alpha = \ln[S(\nu_{1})/S(\nu_{2})] / \ln(\nu_{1}/\nu_{2})$, we
have $\sigma_\alpha = \{[\sigma_{S(\nu_{1})}/S(\nu_{1})]^2 + 
[\sigma_{S(\nu_{2})}/S(\nu_{2})]^2\}^{0.5} / \ln(\nu_{1}/\nu_{2})$.
For spectral indices involving the GISMO (150 GHz) and VLA (1.4 or 5 GHz) data,
the random uncertainties are dominated by the GISMO data. Except at the faint edges
of the SNR, the random uncertainties in the GISMO data are $\lesssim3\%$ 
(See section 2.1). Thus $\sigma_\alpha \lesssim \{\sigma_{S(150)}/S(150)\} / \ln(150/5)
\approx 0.01$. 

Although the random uncertainties spectral index maps are relatively 
modest, there are at least four potential sources of systematic errors which 
may affect the
apparent structure in these spectral index maps. The first is the significant difference 
in the resolution of the original maps. The radio maps were convolved to the 
resolution of the GISMO image (\S \ref{subsec:radiod}). However, if this process 
did not accurately reproduce the GISMO beam, then there would be artificial 
changes in the spectral index in regions where there are steep brightness gradients. 
To check for errors introduced by a resolution mismatch, the convolved 
resolution of the radio map was changed by a factor of 50\%. 
This produced minimal changes the spectral index map, indicating that the 
observed spectral index variations are not a result of mismatched resolution.

A second potential source of error is improper background subtraction. 
The background levels were checked by examining the correlation of the 
radio and 2 mm intensities at intermediate brightness levels 
($100 < I_{{\rm radio}} < 400$ MJy sr$^{-1}$). 
The correlations between the radio and GISMO images are very linear
and extrapolate to zero intercepts indicating 2 mm background errors of
$+2.2$ MJy sr$^{-1}$ with respect to the 21 cm data, and
$-2.2$ MJy sr$^{-1}$ with respect to the 6 cm data.
Comparison with Figure \ref{fig:images} indicates that except in the outermost 
portions of the 
SNR, this background error is $<10\%$ of the observed surface brightness,
and thus any induced errors in the spectral index should be $<0.03$, (much less in 
the brighter central regions). Furthermore, as the background errors have opposite
signs with respect to the 21 and 6 cm images, yet the spectral index maps are very
similar, we conclude that improper background subtraction cannot account for the 
observed structure in the central regions of the spectral index maps. However,
at the edges of the Crab Nebula, the emission of the SNR falls to levels at
or below the uncertainty in the background. Therefore, we cannot reliably 
determine changes in the spectral index at the edges of the SNR and 
do not draw conclusions about the spectral index in the outermost $\sim1'$ 
of the maps shown in Figure \ref{fig:index}.

\begin{figure}[t]
\begin{center}
  \includegraphics[]{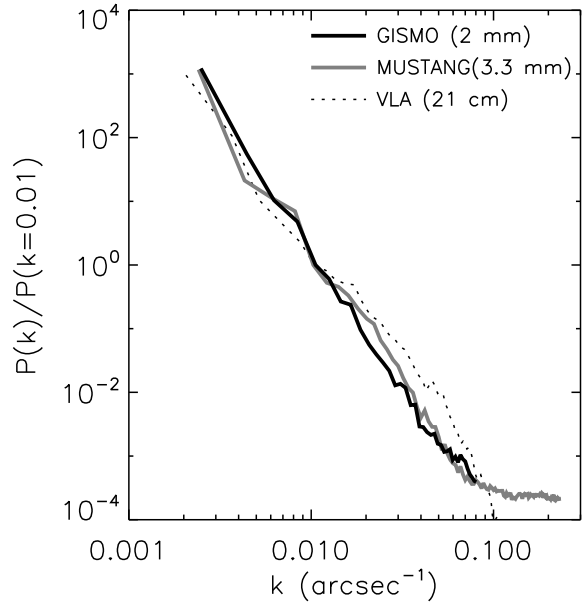}
  \caption{The normalized power spectra of the 2~mm GISMO map, 
  the 3.3 mm MUSTANG map, and the 21~cm (1.4~GHz) radio map, each at its 
  full spatial resolution. At lower spatial frequencies ($k < 0.02$ arcsec$^{-1}$)
  the power spectra are in fair agreement. The larger beams of the GISMO and 
  MUSTANG data cause some attenuation with respect to the VLA data at the higher 
  frequencies. White noise flattens the spectra on scales smaller than the beam 
  (e.g. MUSTANG data). 
   \label{fig:power}}
\end{center}
\end{figure}

A third possible source of error is that either the radio or the 2~mm images may 
be missing flux at particular spatial scales. This may clearly affect the radio data 
due to the limited $u-v$ coverage provide by an interferometer such as the VLA. 
It may also affect the GISMO data due to the periodic nature of the scan pattern
and the processing needed to separate and remove the temporal variation of the 
atmospheric emission.
The radio maps were created using Maximum Entropy Deconvolution provided with a low-
resolution model to account for flux at missing spatial scales. The GISMO maps were 
reduced using the CRUSH package which iteratively removes celestial, foreground, and 
instrument components as measured with the scan pattern one-by-one until all that remains 
is the source and white noise. 
Figure \ref{fig:power} shows the normalized 
spatial power spectra ($P(k)$ as a function of spatial frequency $k$) 
of the 2~mm GISMO map, the 3.3~mm map, and the 21~cm radio map. 
For each data set, the power spectrum is calculated on the images at
their original spatial resolution.
The 2~mm and 21~cm power spectra are well correlated 
at low spatial frequencies ($k < 0.02$, or spatial scales $>50''$). 
Because of the larger PSF, the power seen by GISMO (and MUSTANG) 
becomes increasingly attenuated with respect to that of measured by the VLA 
at smaller scales. At the smallest angular scales, the MUSTANG power 
spectrum flattens due to white noise (instrumental in nature, and thus 
unaffected by the beam). This comparison indicates that 
missing flux at certain spatial scales is not a likely source of the 
observed spectral index variations.

The fourth potential source of error may be caused by the difference 
in the epochs of the two maps. In addition to the overall evolution in the Crab
Nebula's size and brightness, there have been observations of very rapid changes 
at small spatial scales at optical, X-ray, and radio wavelengths \citep[e.g.][]{hester,
greive,bietenholz2004}. However, the similarity between the 21~cm and 6~cm spectral 
index maps provides a strong indicator that the temporal variations are not significant. 
For temporal changes to be responsible for the observed spectral index variations, they 
would need to be negligible between 1987 and 2001 (the epochs of the radio maps), 
and then rapidly become significant in the interval from 2001 to 2008. 

We also attempted to calculate the 2 - 3.3 mm and 1.3 - 2~mm
spectral index maps 
using the GISMO data and either the 3.3~mm MUSTANG data or the 
1.3~mm MAMBO \citep{bandi} observations. However, because 
of the close proximity of the wavelengths, small errors in the background or 
large scale structure lead to magnified effects in the derived spectral index, 
which cause the resulting  spectral index map to be unreliable. 
The mm data sets alone are insufficient to provide
accurate spectral index maps over such small ranges in wavelength.

\section{Discussion}
\label{sec:discussion}

\begin{figure*}[t!]
\begin{center}
  \includegraphics[width=3in]{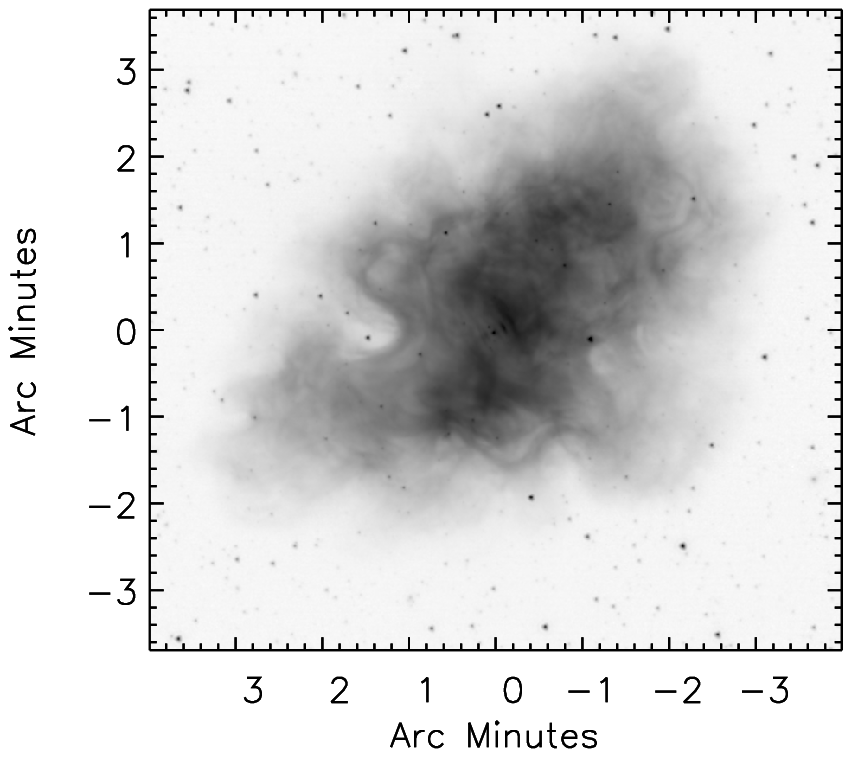}
  \includegraphics[width=3in]{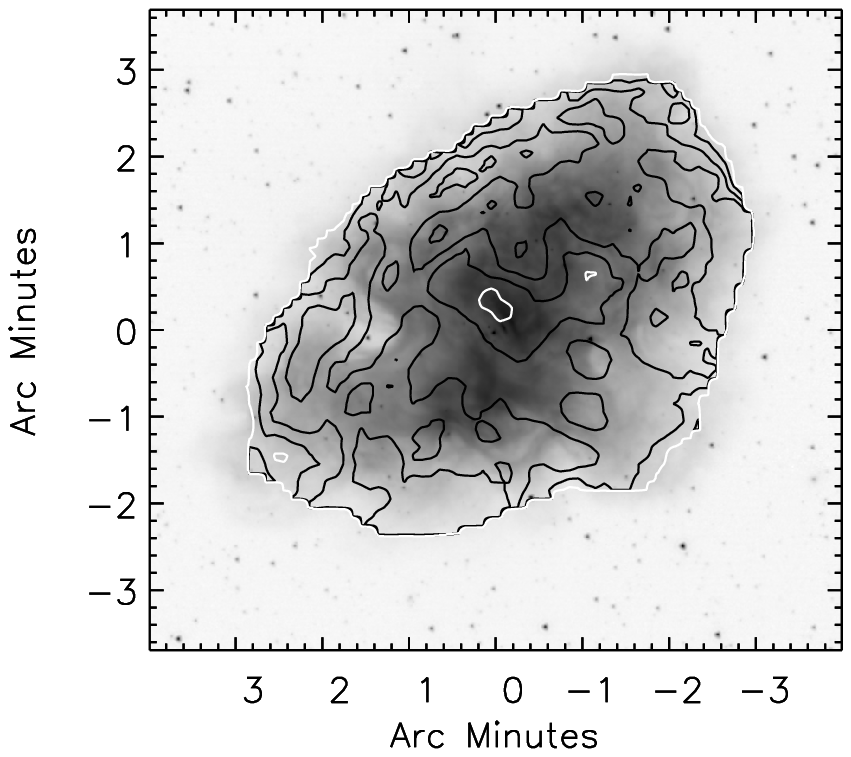}\\
  \includegraphics[width=3in]{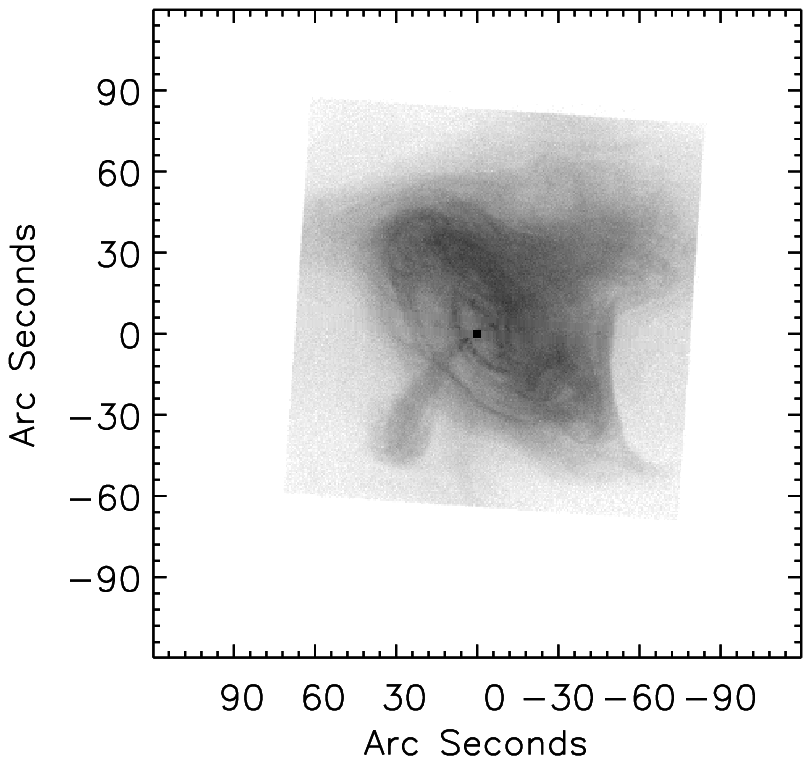}
  \includegraphics[width=3in]{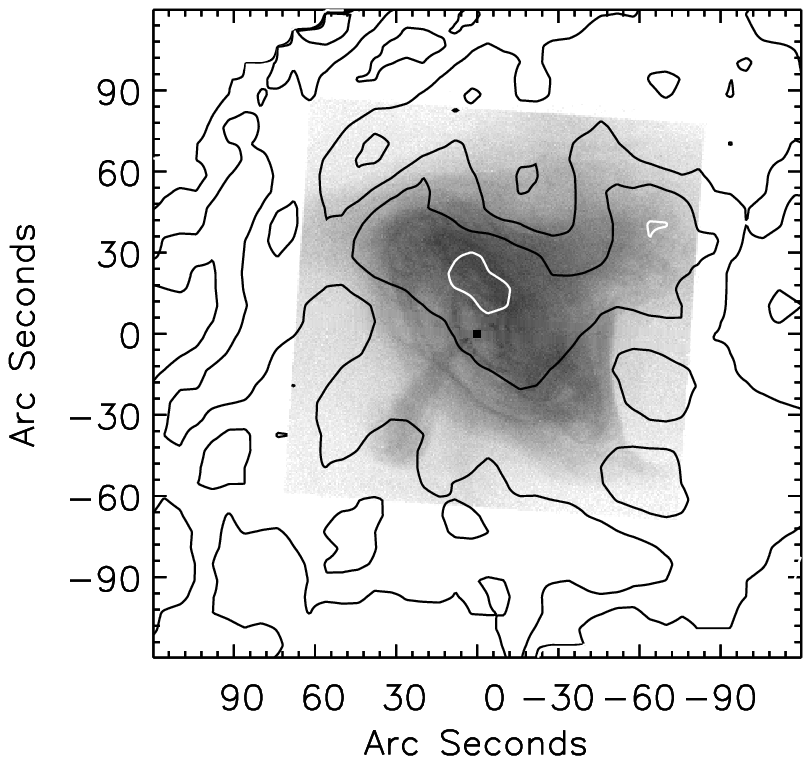}
  \caption{ Comparison of the spectral index variations with synchrotron emission at IR and X-ray wavelengths. 
  (top left) {\it Spitzer} IRAC 4.5 $\micron$ image of the Crab Nebula. Synchrotron 
  emission is dominant at this wavelength. 
  (bottom left) {\it Chandra} ACIS X-ray image of the central portion of the Crab Nebula. 
  The toroidal structure around the pulsar is very clear at this wavelength. 
  (top and bottom right) Spectral index contours $\alpha$, for 
  the 21.0~cm$-$2.0~mm range (from Fig. \ref{fig:index}) overlaid on 
  the left hand images. 
  The areas of flattest spectral index are correlated with the brightest part of the 
  toroidal structure surrounding the Crab Pulsar.  This is especially clear 
  at X-ray wavelengths. 
    The axes indicate right ascension and declination offsets from the pulsar location: 
  $(\alpha_{\rm J2000},\delta_{\rm J2000}) = (05^h\ 34^m\ 32^s,+22\arcdeg\ 00'\ 52'')$. 
\label{fig:overlay}}
\end{center}
\end{figure*}

Having ruled out various sources of systematic error above, we conclude that 
the observed spectral index variations shown in Figure 
\ref{fig:index} are real properties of the SNR, and are not caused by 
systematic effects inherent in the data or the
analysis procedures. The reality of the spectral index variations is 
strengthened by a very strong correlation with those found by 
\cite{bandi} at 1.3 mm and \cite{green} at 850 $\micron$ using different instruments.
Additionally, in Figure \ref{fig:overlay}
we compare the spectral index with the synchrotron emission
as depicted at \dataset[ADS/Sa.Spitzer#0034837248]{IR} and \dataset[ADS/Sa.CXO#obs/01997]{X-ray} wavelengths.
The regions of flatter spectral index are well correlated with the toroidal 
structure of emission that surrounds the pulsar. This structure is present at 
IR wavelengths, but is most distinct at X-ray wavelengths. 

\cite{bandi} have argued that the spectral index variations, along with the integrated 
1.3 mm flux density which appeared to lie above the extrapolation of the radio 
synchrotron spectrum, were indications of a distinct synchrotron component
generated by a separate population of relativistic electrons. With measurements 
at 850 $\micron$, \cite{green} found similar spectral index variations, but with 
better signal to noise and correspondingly less apparent small-scale structure.
However, their integrated flux density is a good match to the extrapolation of
the radio spectrum. They conclude that a second synchrotron component is not required,
and suggest that the known temporal variations in the region near the pulsar may 
be responsible for the flatter spectral index in the region. 
From analysis of WMAP \citep{page}, Archeops \citep{desert, macias-perez2007}, and published integrated flux densities, 
\cite{macias-perez} also conclude that the data are not significantly better fit by the
inclusion of an additional synchrotron or cold dust component to enhance
emission at mm wavelengths.
Our integrated 2 mm flux density measured 
by GISMO supports the conclusions of \cite{macias-perez} and \cite{green}, 
and weaken the argument for a second component. 
However, because we find the same spectral index variations 
when comparing to radio images for two different epochs, we do not believe that 
temporal variation between the times of the radio and mm observations can be 
responsible for the spectral index variations.

The presence of small-scale spectral index variations at wavelengths 
$\lambda < 2$~mm implies the integrated spectrum of the Crab Nebula is
probably better characterized by a more gradual change in slope rather than
a sharp break from one spectral index to another (Eq. 1). 
However, the present far-IR to mm data are insufficient to reveal this distinction
in the integrated emission. The spectral index variations of the synchrotron 
emission also impact study of the dust emission, which is strongest between 
10 and 100 $\mu$m.
Accurate spatial and spectral information for the dust requires subtraction
of the synchrotron component. The synchrotron emission cannot be simply 
extrapolated from longer or shorter wavelengths using the global spectral index.
If local variations are not included, residual synchrotron emission will mistakenly 
be attributed to dust, which will adversely impact assessments of the dust 
temperature, mass, composition, and spatial distribution.

\section{Conclusion}
\label{sec:concl}
We present 2~mm observations of the Crab Nebula using the Goddard IRAM Superconducting 2~mm Observer (GISMO) at the IRAM 30~m telescope. As these are among the first GISMO
observations, they provide a test of the instrument, the observing strategy,
and the data reduction with respect to extended sources.

The GISMO map for the Crab Nebula gives a total integrated flux density of $244\pm24$~Jy which is in accordance with the known radio/sub millimeter spectrum of the Crab Nebula. The radio end of the spectrum is described well using a power law with $\alpha_1 = -0.30 \pm 0.01$ which matches the well-established value of \cite{baars}. 
A simultaneous power law fit to synchrotron emission at 3 -- 8 $\micron$ allows us 
to locate the break in the spectrum to be at $\nu_{b} = 695^{+651}_{-336}$ GHz.
However in detail, the break may not be sharp, and the frequency may vary at 
different locations within the nebula.
We observe no excess of flux which would indicate the presence of a new synchrotron component as proposed in \cite{bandi}. 

Spectral index maps were created to compare the GISMO data with the radio regime. 
Two VLA maps at 1.4 GHz and 5 GHz were used. 
Comparisons with these maps produce similar results and both seem to show 
a trend of flattening of the spectral index in the inner regions of the nebula 
close to the pulsar. The region of flattest spectral index is 
well correlated with the torus in the central region of the nebula as seen at IR 
and X-ray wavelengths. 
We confirm the mm - radio spectral index variations observed by \cite{green} and \cite{bandi},
but as we derive the same spectral index variations when comparing to radio data from 
two different epochs, we conclude that temporal variations in the synchrotron emission 
near the pulsar cannot account for the observed changes in spectral index.

\acknowledgements 
We are extremely grateful to 
M. Bietenholz and R. Bandiera having provided us with the high 
quality radio maps and the 1.3~mm map, respectively, of the Crab Nebula. 
We thank the anonymous referee for suggesting useful improvements to the paper. 
We thank the IRAM 30m telescope support staff in Granada, especially 
Javier Lobato and Dave John, for their assistance on the inaugural 
GISMO observing run.
Support for this work was provided through NSF grants AST-0705185 
and AST-1020981 and NASA grant APRA 04-0055-149. 
The National Radio Astronomy Observatory is a facility of the National Science 
Foundation operated under cooperative agreement by Associated Universities, Inc. 
The IRAC map is based on observations made 
with the Spitzer Space Telescope, which is operated by the Jet Propulsion 
Laboratory, California Institute of Technology under a contract with NASA.
This research has made use of data obtained from the Chandra Data Archive
operated by the Chandra X-ray Center (CXC).

%\bibliography{paper}

\end{document}